\newif\ifEPRINT \EPRINTtrue
\ifx\undefined\ifEPRINT
\documentstyle[aps,pra,preprint,epsf]{revtex}
\else
\documentstyle[aps,pra,multicol,epsf]{revtex}
\fi
\newcommand{\bold}[1]{\mbox{\boldmath $#1$}}    
\newcommand{\r}{{\bold{r}}}                     
\newcommand{\p}{{\bold{p}}}                     
\newcommand{\del}{\partial}                     
\newcommand{\Exp}[1]{{\rm e}^{#1}}              

\newcommand{\getps}[2]{\epsfxsize=#1in\epsfbox{#2}}	
\newcommand{\GETPS}[2]{\getps{#1}{#2}}			

\newcommand{\FIGAps}{%
\begin{figure}
\hbox{\hspace*{-1.2cm}\GETPS{4}{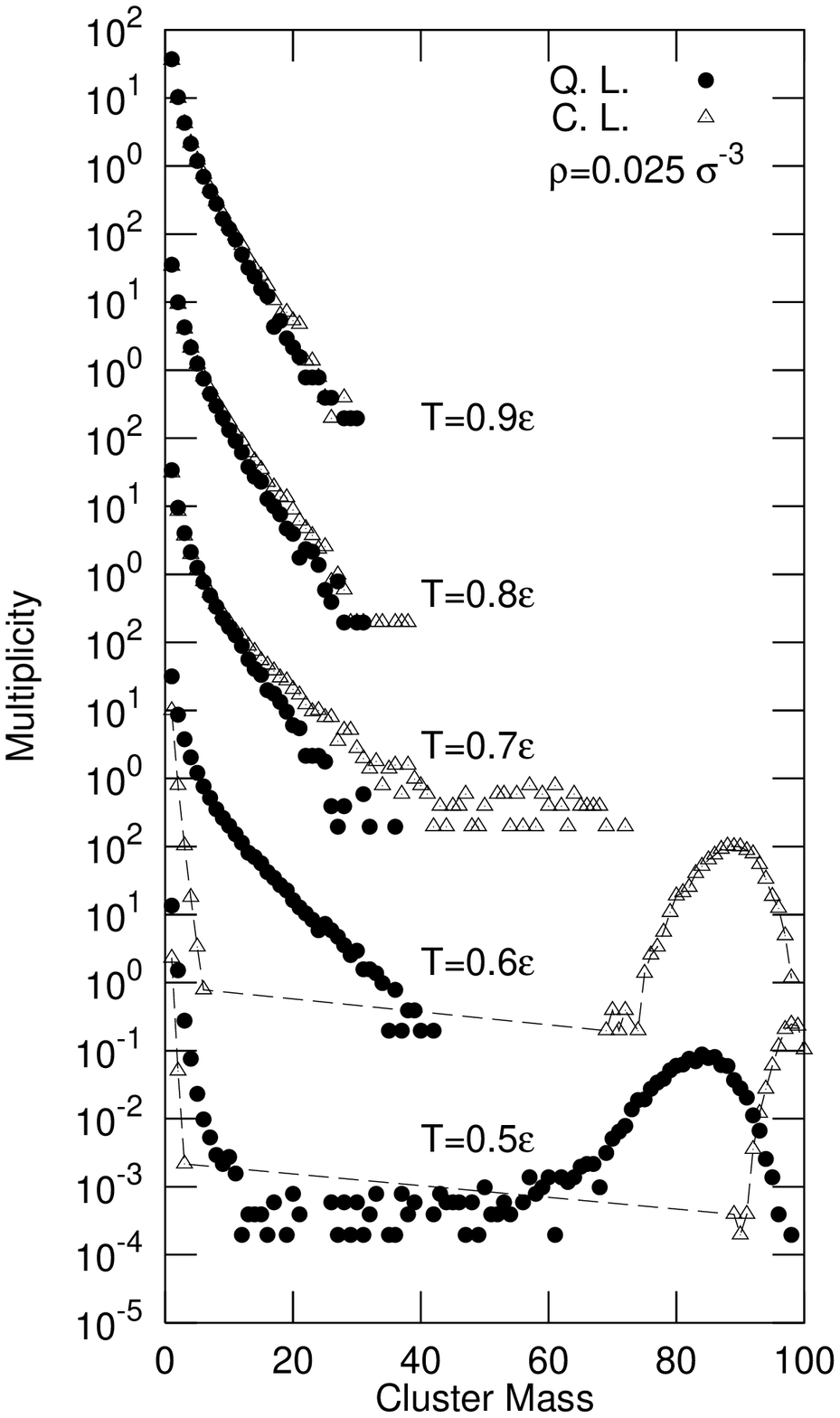}}
\caption{The mass distribution of argon clusters calculated
either {\em with} (solid circle) or {\em without} (open triangle)
the quantum Langevin force emulating the quantum fluctuations
inherent in wave-packet dynamics.
The temperatures are indicated in units of $\epsilon$
and the atomic density is $\rho=0.025\ \sigma^{-3}$.}
\label{fig:1}
\end{figure}
}
\newcommand{\FIGBps}{%
\begin{figure}
\hbox{\hspace*{-1.2cm}\GETPS{4}{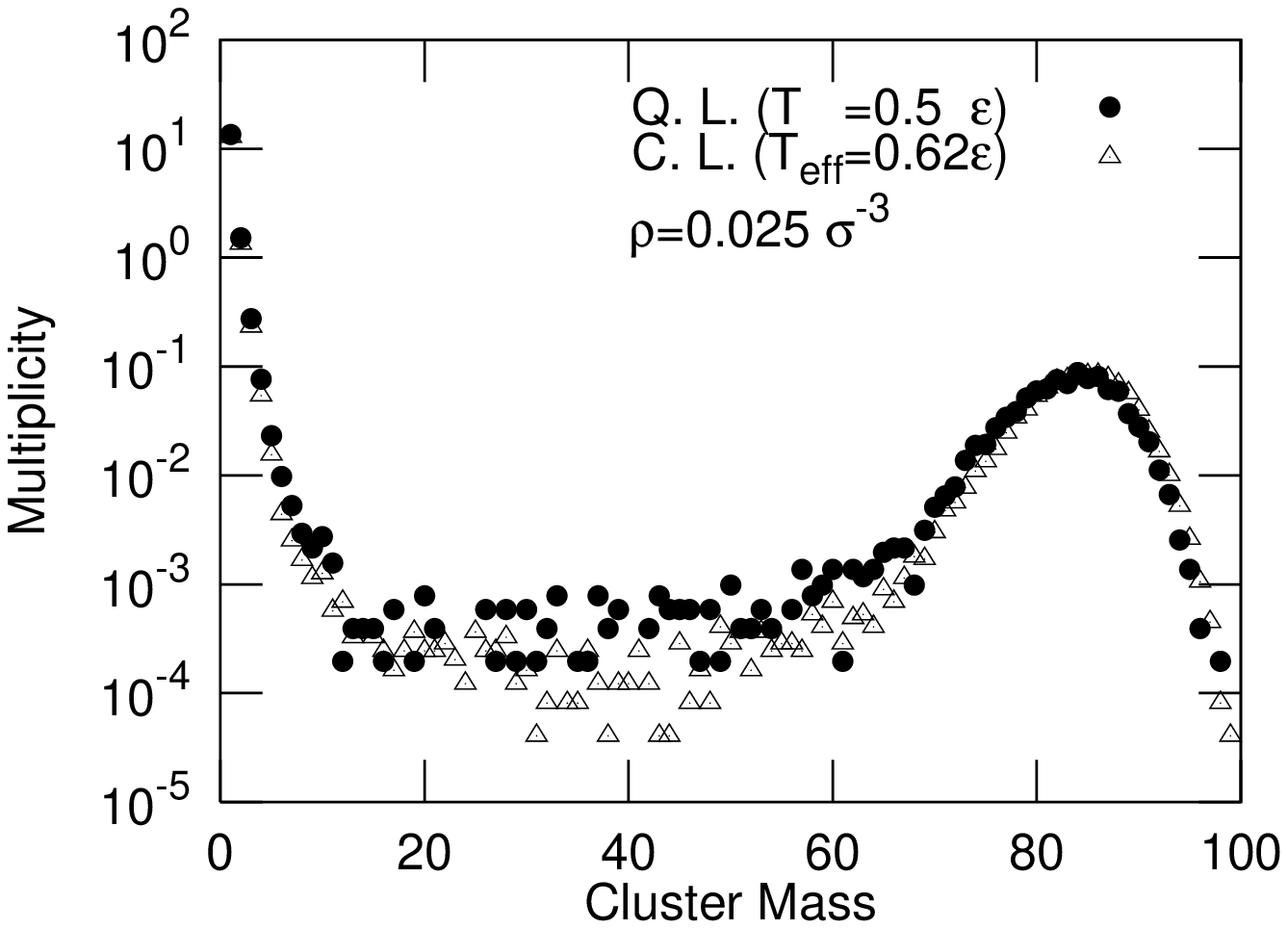}}
\caption{The cluster mass distribution obtained with
either the quantal Langevin model at $T=0.5\epsilon$ (solid circles)
or the classical Langevin model
at the corresponding effective temperature $T_{\rm eff}=0.62\epsilon$
(open triangles),
at the density $\rho=0.025\ \sigma^{-3}$.}
\label{fig:2}
\end{figure}
}

\ifx\undefined\ifEPRINT
\newcommand{\NARROWTEXT}{}
\newcommand{\WIDETEXT}{}
\newcommand{\FIGA}{\begin{center}\fbox{Fig.~\ref{fig:1}}\end{center}}
\newcommand{\FIGB}{\begin{center}\fbox{Fig.~\ref{fig:2}}\end{center}}
\renewcommand{\GETPS}[2]{}
\newcommand{\FIGCAP}{
\FIGAps
\FIGBps
}

\else
\newcommand{\NARROWTEXT}{\begin{multicols}{2} \global\columnwidth20.5pc}
\newcommand{\WIDETEXT}{\end{multicols} \global\columnwidth42.5pc}
\newcommand{\FIGA}{\FIGAps}
\newcommand{\FIGB}{\FIGBps}
\newcommand{\FIGCAP}{}
\fi

\begin{document}
\ifx\undefined\ifEPRINT\begin{titlepage}\thispagestyle{empty}\fi
\preprint{LBL-39754}
\draft

\title{Quantum Fluctuations Affect the Critical Properties of Noble Gases}

\author{Akira Ohnishi$^{a,b}$ and J\o rgen Randrup$^b$}

\address{$^a$ Department of Physics, Faculty of Science,
                Hokkaido University, Sapporo 060, Japan}

\address{$^b$ Nuclear Science Division, Lawrence Berkeley Laboratory,
		University of California, Berkeley, California 94720}
\date{\today}

\maketitle

\begin{abstract}
Interacting argon atoms are simulated
with a recently developed quantum Langevin transport treatment
that takes approximate account of the quantum fluctuations
inherent in microscopic many-body descriptions based on wave packets.
The mass distribution of the atomic clusters
is affected significantly near the critical temperature
and thus it may be important to take account of quantum fluctuations
in molecular-dynamics simulations of cluster formation processes.
\end{abstract}


\ifx\undefined\ifEPRINT\end{titlepage}\fi

\NARROWTEXT

\newcommand{\AO}[1]{{\sl #1}}
\newcommand{\JR}[1]{$<$JR:{\it #1}:JR$>$}

Molecular dynamics presents a powerful tool
for elucidating the features of mesoscopic systems \cite{MD}.
The present communication draws attention to the possible
importance of quantum fluctuations in such treatments.
To illustrate this issue,
we focus on the cluster mass distribution
for argon atoms in thermal equilibrium.

Entities interacting via van der Waals forces,
such as atoms of noble gases,
form clusters exhibiting enhancements at magic mass numbers
associated with especially compact geometric configurations \cite{Farges}.
Several studies based on molecular dynamics
have been devoted to the exploration of the mass spectrum of argon clusters
in various scenarios
(see for example \cite{Vicentini,Blink,Beck,Dumont,Ikeshoji}).
However, the detailed formation process of the clusters,
and their observed relative abundance, is not yet fully understood
and may well depend on the specific experimental conditions.
Because of this delicacy,
we examine here the possible role of quantum fluctuations.
Such fluctuations are in principle present in the the classical description
because it is based on the assumption that
the many-body wave function can be regarded as
a product of single-particle wave packets, which are not energy eigenstates.

Specifically, our discussion invokes a recently developed treatment
in which this inherent effect is emulated by a quantal Langevin force
augmenting the usual classical molecular-dynamics equations of motion%
\cite{OR95,OR96a}.
That treatment was developed in the context of nuclear fragmentation processes
and we first briefly describe its adaptation to atomic systems.
Subsequently we use the model to calculate the cluster mass spectrum
for thermal ensembles of argon atoms
and discuss how the results compare with the usual classical simulation.

The interaction between two atoms $i$ and $j$
is described by the usual Lennard-Jones potential,
\begin{equation}\label{V}
V_{ij}=
4\epsilon\left(({\sigma\over r_{ij}})^{12}-({\sigma\over r_{ij}})^6\right)\ ,
\end{equation}
where the values $\sigma=3.405$ \AA\ and $\epsilon=119.8$ K
are appropriate for argon atoms
and they provide convenient scales for distance and energy, respectively.
(For numerical convenience,
the potential is truncated at $r_{ij}=6\sigma$
which has no effect on the results.)
The minimum in the potential occurs at an interatomic separation of
$r_0=2^{1/6}\sigma\approx1.1225\ \sigma$.
Furthermore,
the strength of the interaction is characterized by the parameter $\epsilon$
which equals the magnitude of the potential at its minimum, $V_0=-\epsilon$,
and provides a convenient energy unit.

The simulation of a classical fluid in thermal equilibrium
can be accomplished by augmenting the molecular-dynamics equations of motion
by a stochastic term representing the coupling to a thermal reservoir,
\begin{eqnarray}
\dot{\r}	&=& {1\over m}\p\ ,\\ \label{eq:CL}
\dot{\p}	&=& \bold{f}\ -\ \nu\bold{M}^2\cdot\p\
+\ \sqrt{\nu T}\ \bold{M}\cdot\bold{\zeta}\ .
\end{eqnarray}
Here $\bold{f}_i=-\sum_j\bold{\nabla}V_{ij}$
is the force exerted on the atom $i$ by all the other atoms $j\neq i$
via the employed Lennard-Jones potential (\ref{V}).
Furthermore,
the Langevin force $\bold{\zeta}$ emulating the coupling to the reservoir
is a standard white noise with the correlation function
\begin{equation}\label{C-L}
\prec\bold{\zeta}_i(t)\bold{\zeta}_j(t')\succ\
=\ 2\delta_{ij}\ \bold{I}\ \delta(t-t')\ ,
\end{equation}
where $\bold{I}$ is the $3\times3$ unit matrix.
The rate $\nu$ determines the strength of the coupling
and should be chosen to be sufficiently small that its value
does not affect the results.
Finally, the matrix $\bold{M}$ has the elements
$M_{ij}=(\delta_{ij}-1/N)\bold{I}$,
and so the Langevin force does not affect the overall momentum
$\bold{P}=\sum_i \p_i$.

When the equations of motion are propagated by a simple leap-frog method,
the increment in the atomic momenta can be calculated as
$\p_i\rightarrow\p_i+\Delta\p_i$ with
\begin{eqnarray}\label{DpCl}\nonumber
\Delta\p_i&=&\bold{f}_i\Delta t\ -\ \nu\Delta t\p_i\
+\ \sqrt{\nu\Delta t T}\bold{\chi}_i\\
&+&{1\over N}
(\nu\Delta t\sum_j\p_j -\sqrt{\nu \Delta t T}\sum_j\bold{\chi}_j)\ ,
\end{eqnarray}
where $\bold{\chi}_i$ are a random vectors whose components have
a normal distribution, $\prec\bold{\chi}_i\succ=\bold{0}$ and
$\prec\bold{\chi}_i\bold{\chi}_j\succ=2\delta_{ij}\bold{I}$.

In the quantal treatment,
each individual atom is described by a Gaussian wave packet
with a spatial variance equal to $\Delta r^2=\hbar/2m\omega$,
where $m$ is the atomic mass and the frequency $\omega$
is determined by from the relation $\omega^2=V_0''/m$.
The corresponding ``level spacing'' is
\begin{equation}\label{D}
D\equiv\hbar\omega=
\hbar c\left({V_0'' \over mc^2}\right)^{1\over2}\approx0.2265\ \epsilon\ .
\end{equation}
The inclusion of the quantum Langevin force
effectively modifies the drift term for the intrinsic cluster motion
by a factor given by \cite{OR96a,OR96b}
\begin{equation}\label{alpha}
\alpha={T\over D} \left(1-{\rm e}^{-D/T}\right) < 1\ .
\end{equation}
In terms of this reduction factor $\alpha$, 
the quantal Langevin equation can be written as follows.
\begin{equation}\label{eq:QL}
\dot{\p} =
	\bold{f}
	-\nu \bold{M}^2\cdot [ \alpha (\p - m \bold{V}) + m\bold{V} ]
	+\sqrt{\nu T}\bold{M}\cdot\bold{\zeta}\ , 
\end{equation}
where $\bold{V}$ is the velocity of the cluster to which the atom belongs.
The only complication arises from the requirement that
the center-of-mass motion of each individual cluster
retain a classical character.
Thus, in order to determine $\bold{V}$,
it is necessary to perform a cluster analysis
of the $N$-body system at each time step.
This task can be easily accomplished
by considering two atoms as belonging to the same cluster
if (and only if) their separation is less than a certain critical value,
taken to be equal to $2\sigma$.
When the drift term is modified correspondingly,
the expression (\ref{DpCl}) for the momentum increment
is being replaced by
\begin{eqnarray}\label{DpQu}\nonumber
\Delta\p_i&=&\bold{f}_i\Delta t
-\nu\Delta t [ \alpha (\p_i -m\bold{V}_n) + m\bold{V}_n ]
+\sqrt{\nu\Delta t T}\bold{\chi}_i\\
&+&{1\over N}
(\nu\Delta t\sum_j\p_j -\sqrt{\nu \Delta t T}\sum_j\bold{\chi}_j)\ ,
\end{eqnarray}
where $\bold{V}_n$ is the velocity of the cluster $n$ containing $i$.

The occurrence of the quantum Langevin force
is a consequence of the fact that the wave packets are not energy eigenstates.
That same feature causes another complication,
namely the need for taking account of the thermal distortion
of the intrinsic structure of each wave packet
(the relative weight of each energy component of the wave packet
is temperature dependent).
This feature can be taken into account by subjecting the dynamical state
of the system to a cooling process
before making the observation\cite{OR96a,OR96b}.
In the present case,
this amounts to integrating the equation
\begin{equation}
\label{eq:Cool}
{\del\p_i\over\del\tau}=-{2\Delta p^2\over\hbar}(\p_i - m\bold{V}_n)\ ,\
{\del\r_i\over\del\tau}= {2\Delta r^2\over\hbar}\bold{f}_i
\end{equation}
from $\tau=0$ to $\tau=\hbar/2T$,
thereby distorting the current dynamical values $(\r(t),\p(t))$
into $(\r'(t),\p'(t))$.
As a result of this cooling process,
the atoms change their positions
and so the cluster structure may be modified.
While this is a significant effect in the case of nuclear fragmentation
\cite{OR96b},
the effect is unimportant for atomic systems,
since the width of the wave packet $\Delta r$ is quite small
(less than 5\% of $\sigma$ in the case of argon).

\FIGA

In order to illustrate the effect of the energy fluctuations
caused by the use of wave packets,
we show in Fig.\ \ref{fig:1} the cluster mass distributions
obtained for a system of argon atoms held at a specified temperature $T$.
These calculations consider 100 atoms confined within a cubic box
and subjected to periodic boundary conditions.
The temperatures considered range from well above to below critical.
At high temperatures
the cluster mass distribution falls off rapidly,
whereas there is a preference for condensation into large clusters
at low temperatures.
When the classical Langevin equation (\ref{eq:CL}) is employed,
the critical temperature associated with this gas-liquid phase transition
is approximately $T^{\rm cl}\approx0.6\epsilon$,
as can be seen from the figure.
The inclusion of the quantum Langevin force, Eq.\ (\ref{eq:QL}),
produces a steeper slope in the mass distribution at high temperatures
and the phase transition occurs at a lower temperature,
$T^{\rm qu}\approx0.5\epsilon$,
as might then be expected.

The change of the critical temperature can be understood quantitatively
by noting that it is possible to extract an effective classical temperature
from the quantal Langevin equation (\ref{eq:QL}).
The effective temperature can be estimated by means of the Einstein relation
as the square of the diffusion coefficient divided by the drift coefficient,
\begin{equation}\label{Teff}
T_{\rm eff}	= {\nu T \over \alpha \nu} = {T\over \alpha} 
=D/(1-\Exp{-D/T}) > T\ .
\end{equation}
This estimate ignores the correction terms arising
from the cluster center-of-mass motion, 
as is justified when we are interested in the cluster mass distribution, 
because this observable depends mainly on the intrinsic degrees of freedom
of the cluster.
In addition, 
the cluster mass distribution is not affected much by the cooling process,
as already mentioned.
Thus, it is expected that a calculation with the classical Langevin model
at the effective temperature, $T_{\rm eff}$,
will yield results that are similar to those obtained
with the quantal Langevin model at the real temperature, $T$.

\FIGB

This expectation is indeed borne out,
as shown in Fig.\ \ref{fig:2} 
where we compare the cluster mass distribution
obtained with the quantal model at $T=0.5\epsilon$ 
to the result of the classical treatment carried out
at the corresponding effective temperature $T_{\rm eff}=0.62\epsilon$.
The quantitative similarity between the two distributions is remarkable
and supports the above discussion.

The temperature shift depends the system under consideration
and the specific observables extracted.
For example, the temperature shift is in the opposite direction
for particles in a harmonic oscillator potential\cite{OR96a,OR93}
and for nuclear fragmentation\cite{OR96b}.
This is because the cooling process affects the observables as well as
the modification of the drift term for these systems.
In order to illustrate this feature,
we consider here the evolution of the distorted momentum
({\em i.e.}\ the solution to Eq.\ (\ref{eq:Cool})) which is given by
\begin{equation}
\p'_i(t)\ \equiv\ \p_i(t,\tau={\hbar\over2T})\ =\ \Exp{-D/2T}\ \p_i(t) \ .
\end{equation}
Thus, in the rest frame of the cluster,
the distorted momenta of the constituent atoms
are governed by a modified Langevin equation,
\begin{equation}
\label{eq:QLp}
\dot{\p'} =
	\Exp{-D/2T} \bold{f}
	-\alpha \nu \bold{M}^2\cdot\p'
	+\Exp{-D/2T} \sqrt{\nu T}\bold{M}\cdot\bold{\zeta}\ .
\end{equation}
Since this equation of motion has a classical form
({\em c.f.}\ Eq.\ (\ref{eq:CL})),
we can again invoke the Einstein relation
and extract an effective temperature for the intrinsic cluster motion,
\begin{equation} 
T'_{\rm eff}\ =\ { \Exp{-D/T} \nu T\over \alpha\nu}\ 
=\ D/(\Exp{D/T}-1) < T \ .
\end{equation}
It has been shown that calculations with classical molecular dynamics
at this equivalent temperature $T'_{\rm eff}$
yields results that are very similar to the exact quantal results
for the real temperature $T$,
for non-interacting particles in a harmonic potential \cite{OR96a,OR93,SF,OH}.
It is to be expected that this feature holds as well
for the distribution of intrinsic excitation energies of atomic clusters.
Therefore, it would be interesting to measure the atomic cluster mass 
distribution simultaneously with the intrinsic excitation energy distribution,
since the associated temperature shifts are predicted to be of opposite sign
(and thus they could not be mocked up by a simple readjustment of the
classical temperature).

The present treatment takes account of the fact that
the intrinsic atomic motion in a cluster and its center-of-mass motion
are of different character.
While the latter can be considered as classical,
the quantal features of the former may be significant
and lead to modifications of the Einstein relation,
as we have discussed in the framework of wave-packet dynamics.
This important point was already noted by
Ikeshoji et al.~\cite{Ikeshoji}
in their recent study of magic-cluster formation
and it might thus be interesting to apply the present treatment 
to the cluster formation process in an expanding and cooling noble gas.

In the present paper we have adapted a recently developed quantal Langevin
treatment to a system of argon atoms in thermal equilibrium.
The method was developed in the context of nuclear dynamics
and takes approximate account of the energy fluctuations
that are necessarily present when wave packets are used to describe
the system.
The presence of these quantum fluctuations changes the character
of the specific heat from classical to quantal
and their inclusion by the developed method
leads to a significant improvement of the statistical properties
in a number of simple test cases that can be subjected to exact analysis
\cite{OR96a,OR93}
as well as for finite nuclei\cite{OR95,OR96a}.
When incorporated into microscopic dynamical simulations
of nuclear collisions,
it leads to a significant improvement of the calculated fragment
mass distribution \cite{OR96b}.
For atomic systems,
the effect of quantum fluctuations is generally expected to be small,
due to the relatively large mass of the ``particles''.
Nevertheless, as we have shown above,
the inherent quantum fluctuations may also play a role in atomic physics.
Certainly, as our study brings out,
it appears that the critical properties of noble gases
are affected significantly.
The quantum fluctuations may therefore also affect the formation process
and might affect the outcome of dynamical simulations
aimed at understanding the observed mass distributions.\\

This work was supported in part by
the Grant-in-Aid for Scientific Research (No.\ 06740193)
from the Ministry of Education, Science and Culture, Japan,
and by the Director, Office of Energy Research,
Office of High Energy and Nuclear Physics,
Nuclear Physics Division of the U.S. Department of Energy
under Contract No.\ DE-AC03-76SF00098.
One of the authors (AO) also thanks the Ministry of Education,
Science and Culture, Japan, for the Overseas Research Fellowship granted.

\ifx\undefined\ifEPRINT \newpage \fi


\ifx\undefined\ifEPRINT \newpage \fi
\WIDETEXT

\FIGCAP
\end{document}